\documentclass[3p,times]{elsarticle}
\pdfoutput=1
\usepackage{ecrc}
\usepackage{graphicx}
\usepackage{booktabs}
\usepackage{subfigure}

\makeatletter
\newcommand\figcaption{\def\@captype{figure}\caption}
\newcommand\tabcaption{\def\@captype{table}\caption}
\makeatother 

\volume{00}

\firstpage{1}

\journalname{Physics Procedia}

\runauth{J.A. Aguilar et al.}

\jid{phpro}

\jnltitlelogo{Physics Procedia}

\usepackage{amssymb}

\usepackage[figuresright]{rotating}

\begin{document}

\begin{frontmatter}



\dochead{}

\title{Luminometer for the future International Linear Collider - simulation
  and beam test results}


\author[agh,ifj]{J.A. Aguilar}
\author[agh]{S.Kulis}
\author[ifj]{W. Wierba}
\author[ifj]{L. Zawiejski}
\author[ifj]{E. Kielar}
\author[ifj]{M. Chrzaszcz}
\author[desy]{O. Novgorodova}
\author[desy]{H. Henschel}
\author[desy]{W. Lohmann}
\author[desy]{S. Schuwalow}
\author[desy,minsk]{K. Afanaciev}
\author[desy]{A. Ignatenko}
\author[desy,btu]{S. Kollowa}
\author[tau]{I. Levy}
\author[agh]{M. Idzik}
\author[ifj]{J. Kotula}
\author[ifj]{A. Moszczynski}
\author[ifj]{K. Oliwa}
\author[ifj]{B. Pawlik}
\author[ifj]{W. Daniluk}

\address[agh]{AGH University of Science and Technology, Dept.\ of Physics
  and Applied Computer Science, ul. Reymonta 19, Krakow 30-059, Poland}
\address[ifj]{Institute of Nuclear Physics PAN, Division of Particle Physics and Astrophysics, ul. Radzikowskiego 152, Krakow 31-342, Poland}
\address[desy]{DESY, Platanenallee 6, D-15738 Zeuthen, Germany}
\address[tau]{Tel Aviv University, Raymond and Beverly Sackler School of Physics and Astronomy, Tel Aviv 69978, Israel}
\address[btu]{Brandenburg University of Technology, D-03013 Cottbus, Postfach 101344, Germany}
\address[minsk]{NCPHEP Minsk, Bogdanova uliza 153, Minsk, Belarus}

\begin{abstract}
LumiCal will be the luminosity calorimeter for the proposed International Large Detector of the International Linear Collider (ILC). The ILC physics program requires the integrated luminosity to be measured with a relative precision on the order of 10e-3, or 10e-4 when running in GigaZ mode. Luminosity will be determined by counting Bhabha scattering events coincident in the two calorimeter modules placed symmetrically on opposite sides of the interaction point. To meet these goals, the energy resolution of the calorimeter must be better than 1.5\% at high energies. LumiCal has been  designed as a 30-layer sampling calorimeter with tungsten as the passive material and silicon as the active material. Monte Carlo simulation using the Geant4 software framework has been used to identify design elements which adversely impact energy resolution and correct for them without loss of statistics. BeamCal, covering polar angles smaller than LumiCal, will serve for beam tuning, luminosity optimisation and high energy electron detection. Secondly, prototypes of the sensors and electronics for both detectors have been evaluated during beam tests, the results of which are also presented here.
\end{abstract}

\begin{keyword}
LumiCal \sep BeamCal \sep forward calorimetry \sep ILC


\end{keyword}

\end{frontmatter}


\section{Forward calorimetry at ILC}
\label{sec:fcal_at_ilc}

\begin{figure}[h!]
  \begin{minipage}[b]{0.5\linewidth}
    \centering
    \includegraphics[width=0.7\textwidth]{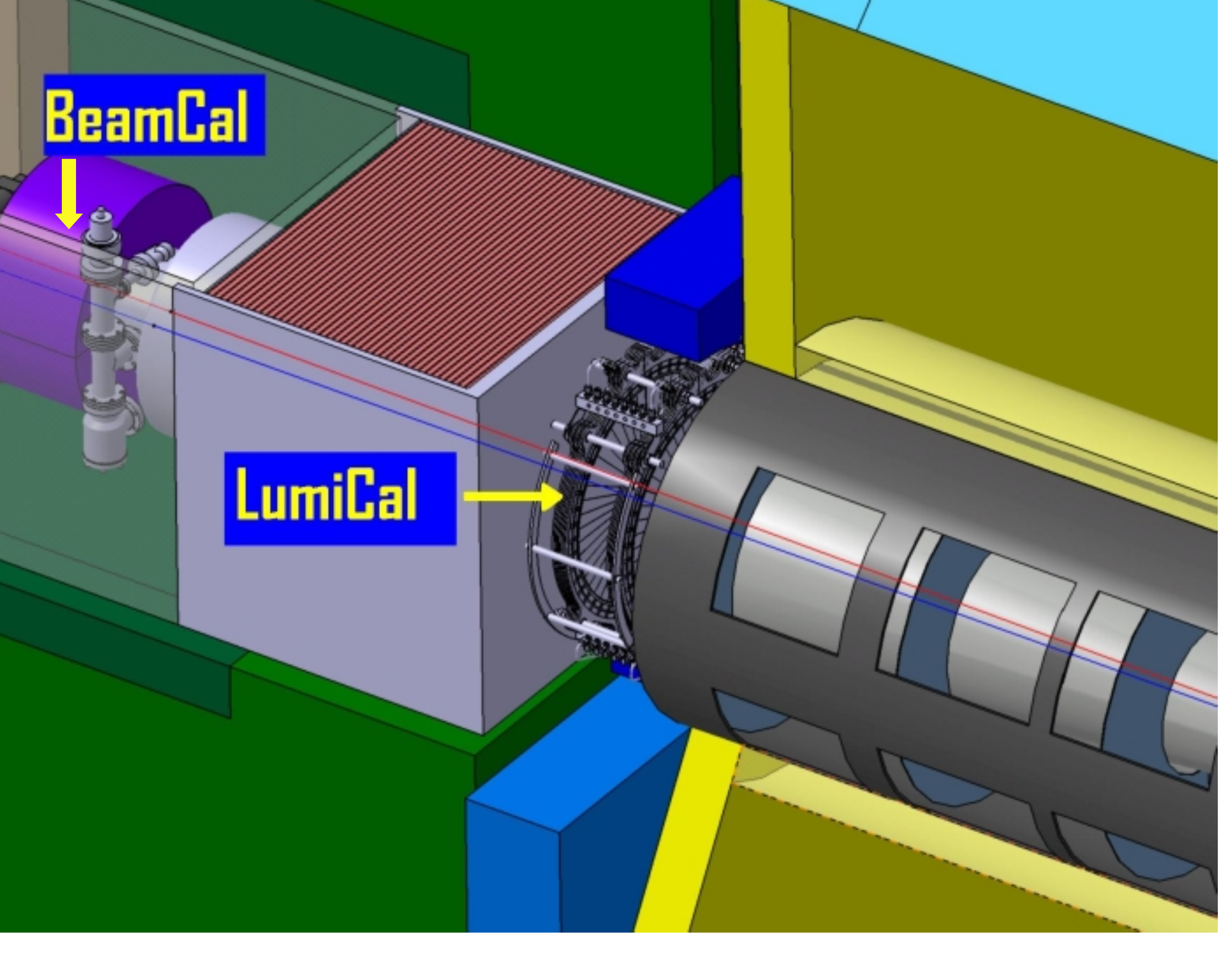}
    \caption{LumiCal and BeamCal in the ILC forward region.}
    \label{fig:ild_schem}
  \end{minipage}
  \begin{minipage}[b]{0.5\linewidth}
    \centering
    \includegraphics[width=0.7\textwidth]{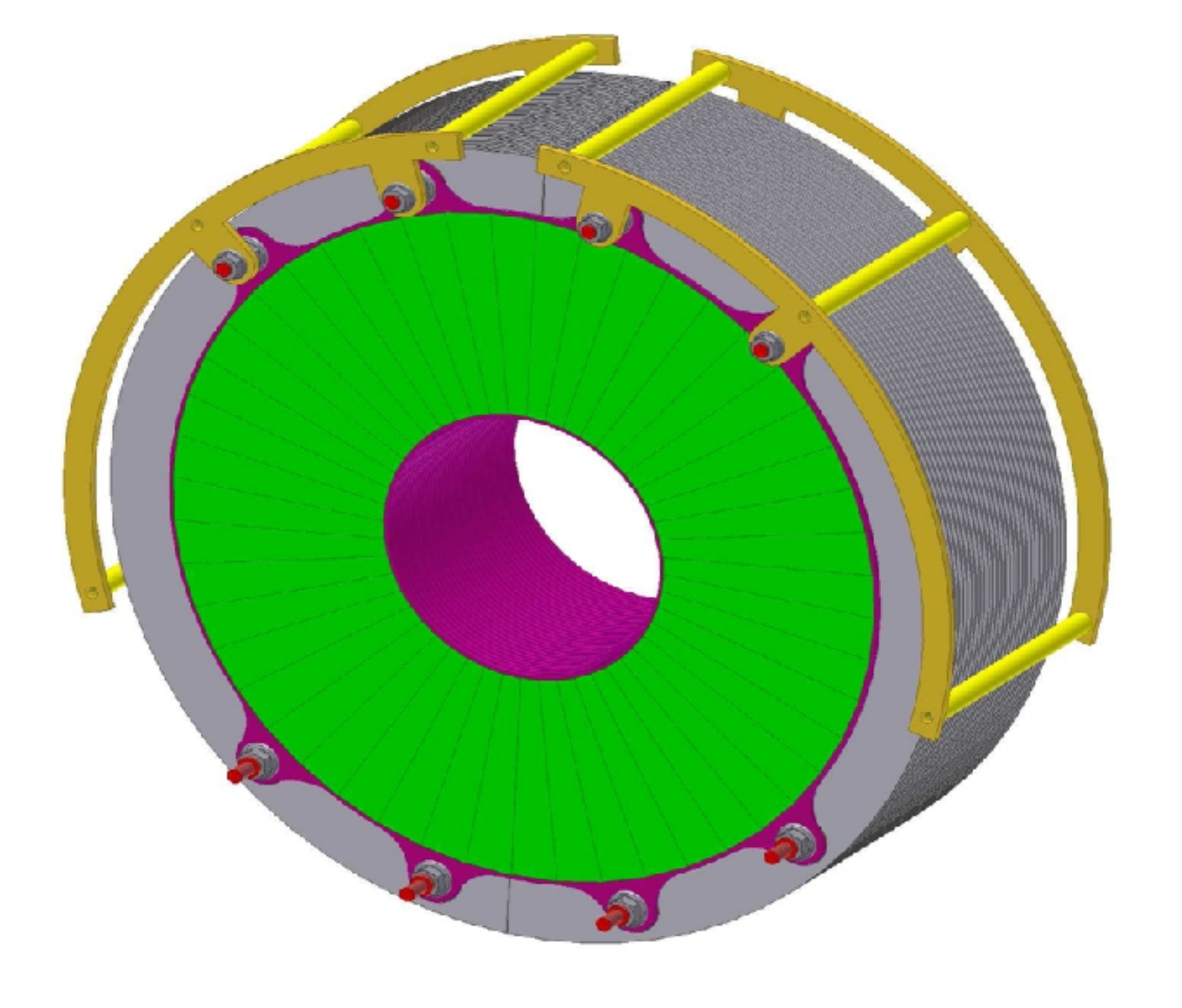}
    \caption{Mechanical model of LumiCal}
    \label{fig:lumical_mech}
  \end{minipage}
\end{figure}

LumiCal is the proposed design for the luminosity calorimeter in the International Large Detector (ILD) \cite{ild_ref} concept for the future International Linear Collider. It consists of two identical detectors placed symmetrically 2.5~m from the interaction point, and covers the angular range from 40~mrad to 69~mrad. BeamCal is proposed for the beam calorimeter, which will sit behind LumiCal at 3.45~m from the IP\@. BeamCal is described in detail in \cite{beamcal_description}. The locations of LumiCal and BeamCal in the ILD are shown in Figure~\ref{fig:ild_schem}. 

\section{LumiCal simulation}
\label{sec:lumical_sim}

\subsection{LumiCal description}
\label{ssec:lumical_geom}
\begin{figure}[h!]
  \begin{minipage}[b]{0.5\linewidth}
    \centering
    \includegraphics[width=0.7\textwidth]{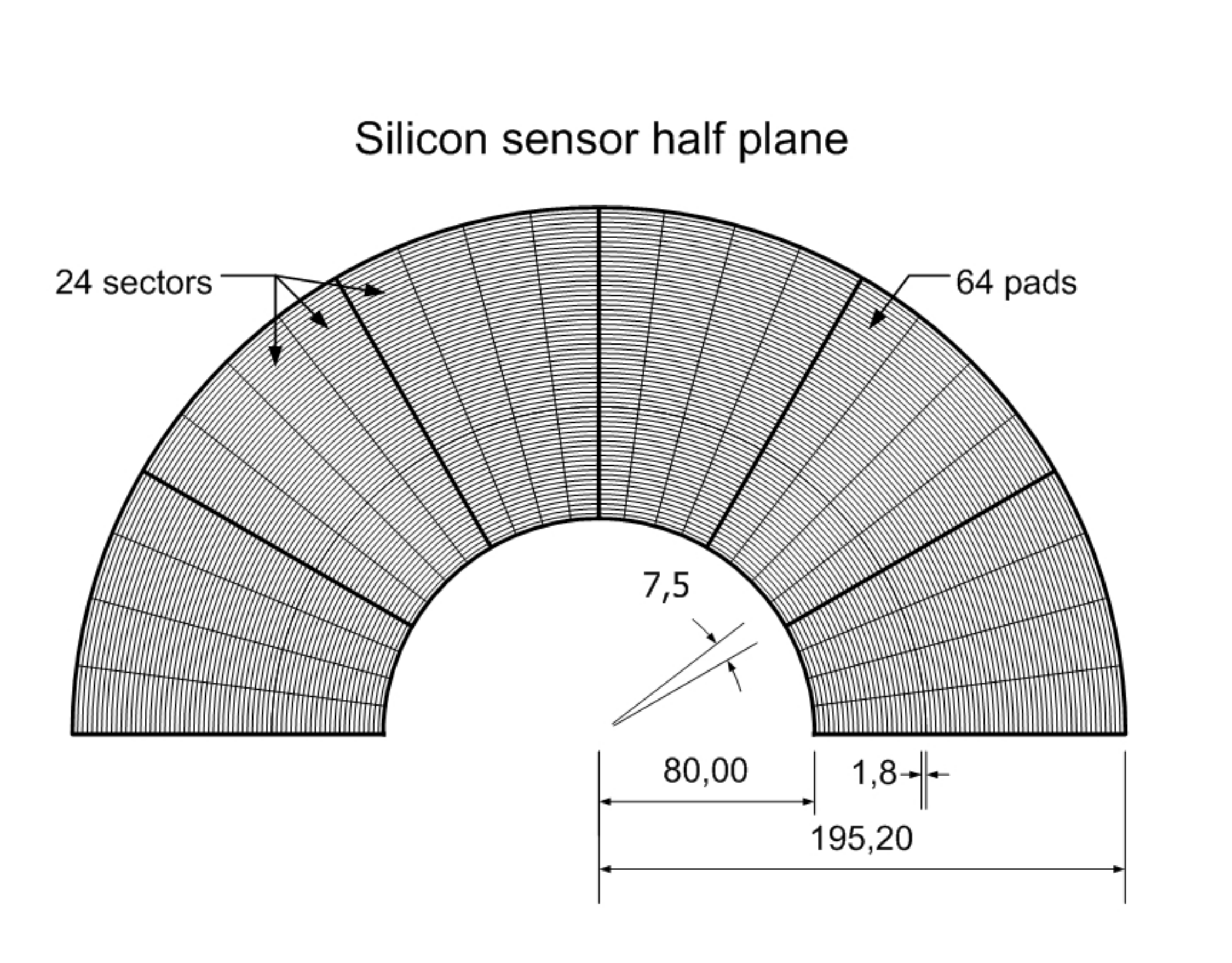}
    \caption{LumiCal sensor half-plane schematic.}
    \label{fig:lumical_sensor_schem}
  \end{minipage}
  \begin{minipage}[b]{0.5\linewidth}
    \centering
    \includegraphics[width=0.7\textwidth]{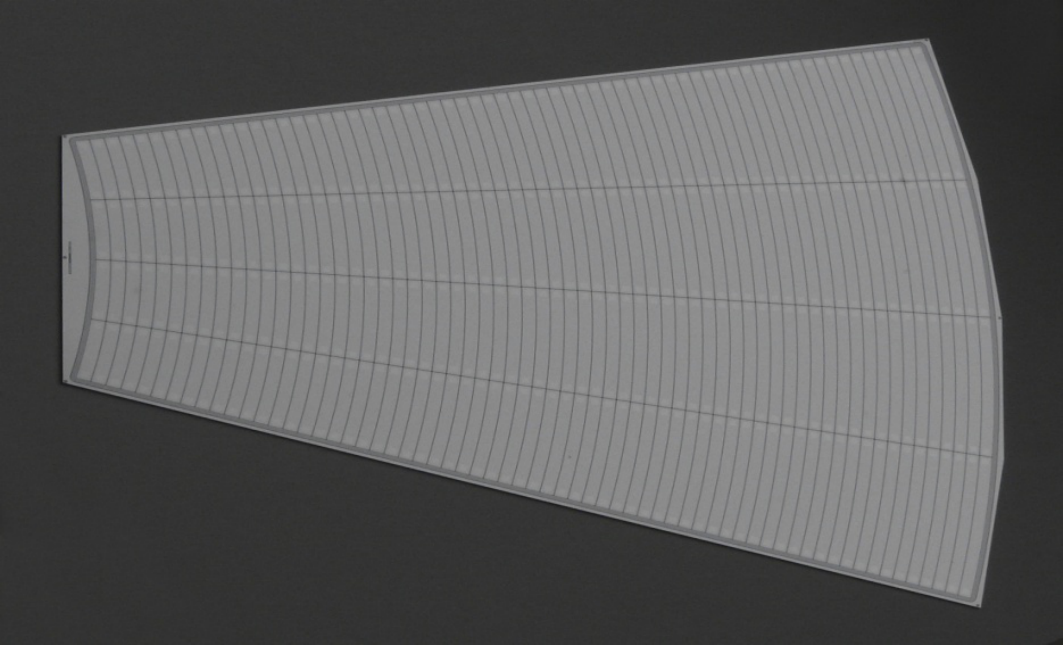}
    \caption{Photograph of LumiCal sensors.}
    \label{fig:lumical_sensors}
  \end{minipage}
\end{figure}

The simulated model of LumiCal was built according to the reference design as
described in \cite{lumical_mech}. A CAD drawing is
shown in Figure~\ref{fig:lumical_mech}. Each LumiCal module will consist of 30
layers of tungsten absorber with silicon sensors and electronics
attached. There will also be space for mechanical support and cooling. The
tungsten will extend radially from 76~mm for the inner edge to 280 mm for the
outer (tentative). The sensitive radius will extend from 80~mm to 195.2~mm, with the rest of
the space on the outside reserved for electronics and support. The sensor
region is divided azimuthally into twelve equal tiles, each covering
30$^{o}$. Between each tile, there will be an uninstrumented 2.4~mm gap. Each
tile is further divided azimuthally into four sectors, each covering
7.5$^{o}$, and radially into 64 pads with a 1.8~mm pitch. In order to
fit around the beam pipe, the LumiCal modules will be built in two
halves and then connected. A schematic of a half-plane of sensors is
shown in Figure~\ref{fig:lumical_sensor_schem}. The sensors are
described in section~\ref{ssec:readout_chain}. A prototype sensor tile
is shown in Figure~\ref{fig:lumical_sensors}.

\subsection{Simulation parameters}
\label{ssec:lumical_sim_params}
The LumiCal simulation code, LuCaS, was built in Geant4 \cite{geant4-1, geant4-2}, and the analysis of simulated events is performed using ROOT \cite{root}. The geometric parameters are maintained to be identical to those in the collaborative ILC software framework \cite{ilcsoft}, but LuCaS has the advantage of being more portable. This allowed it to be ported to the Cyfronet cluster computing facilities \cite{cyfronet}. LumiCal has been studied extensively in simulation previously \cite{iftach_msc}, but the energy resolution and energy scale uncertainty have only been studied with an idealized model with no gaps between sensor tiles (as described in section~\ref{ssec:lumical_geom}).

The simulations tested the response of LumiCal to single high-energy
electrons between 5 GeV and 500 GeV, spread uniformly over the
surface area of the detector. The simulation parameters are given in
table~\ref{tab:sim_params}.

\begin{minipage}[t]{1.0\linewidth}
  \begin{minipage}[c]{0.4\linewidth}
      \centering
      \includegraphics[width=0.8\textwidth]{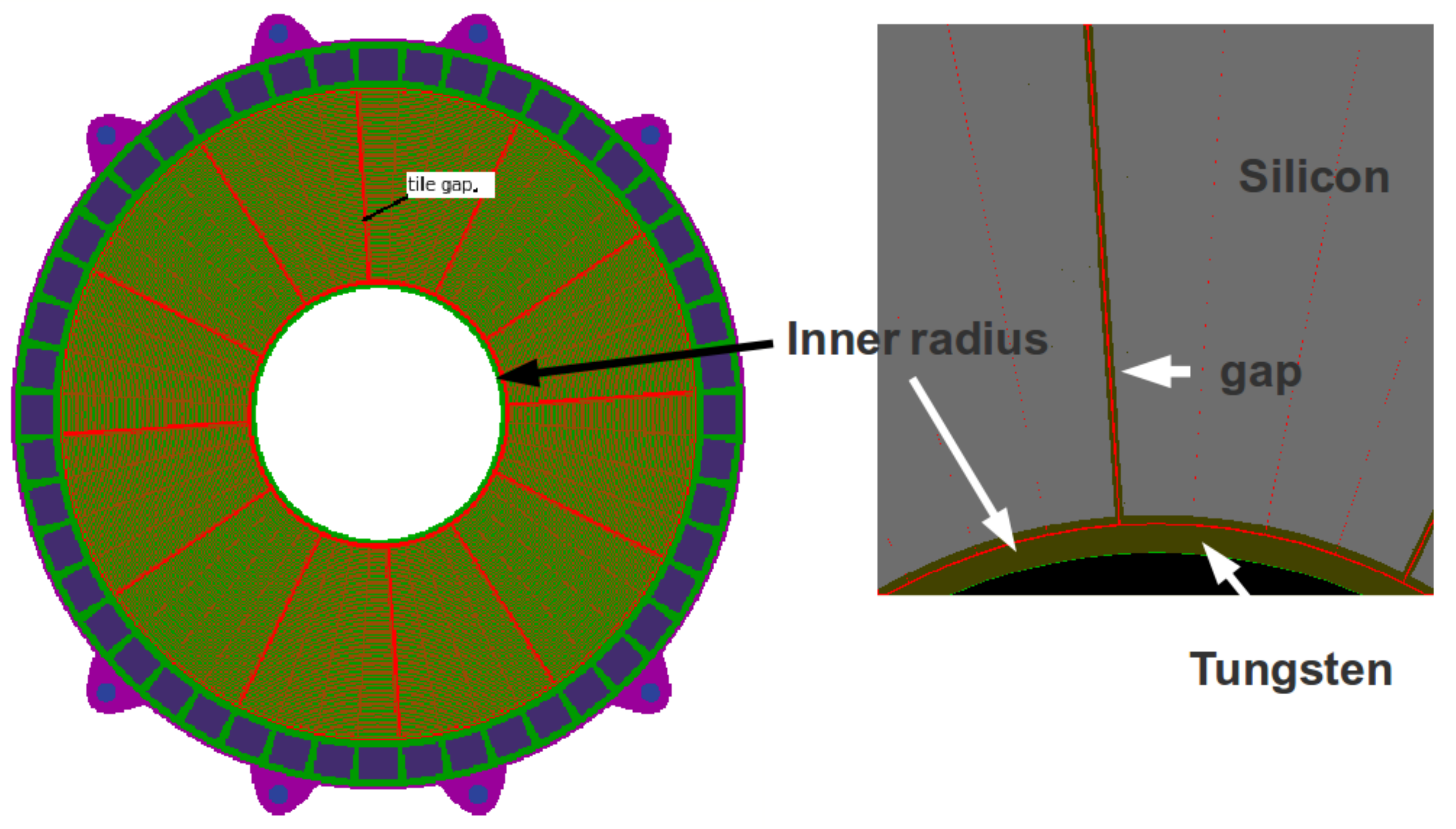}
    \figcaption{Simulated LumiCal model}
    \label{fig:lumical_sim_model}
  \end{minipage}
  \begin{minipage}[c]{0.55\linewidth}
      \centering
      \tabcaption{Simulation parameters}
      \label{tab:sim_params}
      \begin{tabular*}{\textwidth}{l l}
        \toprule
        Parameter & Value \\
        \midrule
        Particles & Single e- \\
        Azimuthal range [rad]& $\phi \in [0, 2\pi]$ \\
        Polar angle range [mrad]& $\theta \in [0.033, 0.073]$\\
        Energies [GeV] & 5, 25, 50, 100, 150, 200, \\
        & 250, 500 \\
        Events/energy & 5000 \\
        Physics lists & ILC Physics (LCPhysicsList) \\
        Range cut & 5 $\mu m$ \\
        \bottomrule
      \end{tabular*}
  \end{minipage}
\vspace{0.3cm}
\end{minipage}
Lastly, three versions of the LumiCal geometry were simulated: the reference design as described previously, and two variants. The first variant did not have alternate layers rotated by 3.75$^o$, so all the tile gaps were aligned directly. The purpose of this model was to make correction simpler at the expense of greater leakage in the gaps. The second variant was an idealized model of LumiCal with no dead areas in between tiles.
 
\subsection{Tile gap effect}
\label{ssec:tile_gap}
The tile gaps themselves cover approximately 3.5\% of the total surface area
of LumiCal. However, their influence extends further due to the Moliere radius
of LumiCal, which is about 15~mm. Thus, 20\% of showers may deposit
some of their energy in the tile gaps, where it will not be
recorded. This produces a long left tail of the raw total energy
deposition per particle, as compared to ideal models of LumiCal
(Figure~\ref{fig:evis_norot}). This is significant because it
increases the RMS of the distribution of visible energy and makes
reconstruction of the original energy more uncertain. It would be
useful to have a method of correcting for the geometric irregularities
that cause this effect, in order to reconstruct the energies of
particles in the gaps as accurately as for particles in the sensors.

Previously, energy resolution was modeled using a single-parameter fit proportional to stochastic fluctuations in the shower, but a term can also be included to account for energy missed by the calorimeter: 

\begin{equation}
  \label{eq:eres}
  \frac{\sigma_{E}}{E} = \sqrt{\frac{a^{2}}{E} + b^{2}},
\end{equation}

where \emph{a} is the stochastic parameter and \emph{b} is the constant
parameter. Further terms can be added to take into account other sources of noise, but were ignored since those sources were not included in the simulation.

The energy resolution for particles of a given energy is calculated
from simulation by dividing the RMS of the reconstructed energy by the
initial energy. This is performed for different energies (here, eight)
and the results are compared with equation~\ref{eq:eres}.

\subsection{Correcting for the tile gaps}
\begin{figure}[h!]
  \begin{minipage}[t]{0.5\linewidth}
    \centering
    \includegraphics[width=0.7\textwidth]{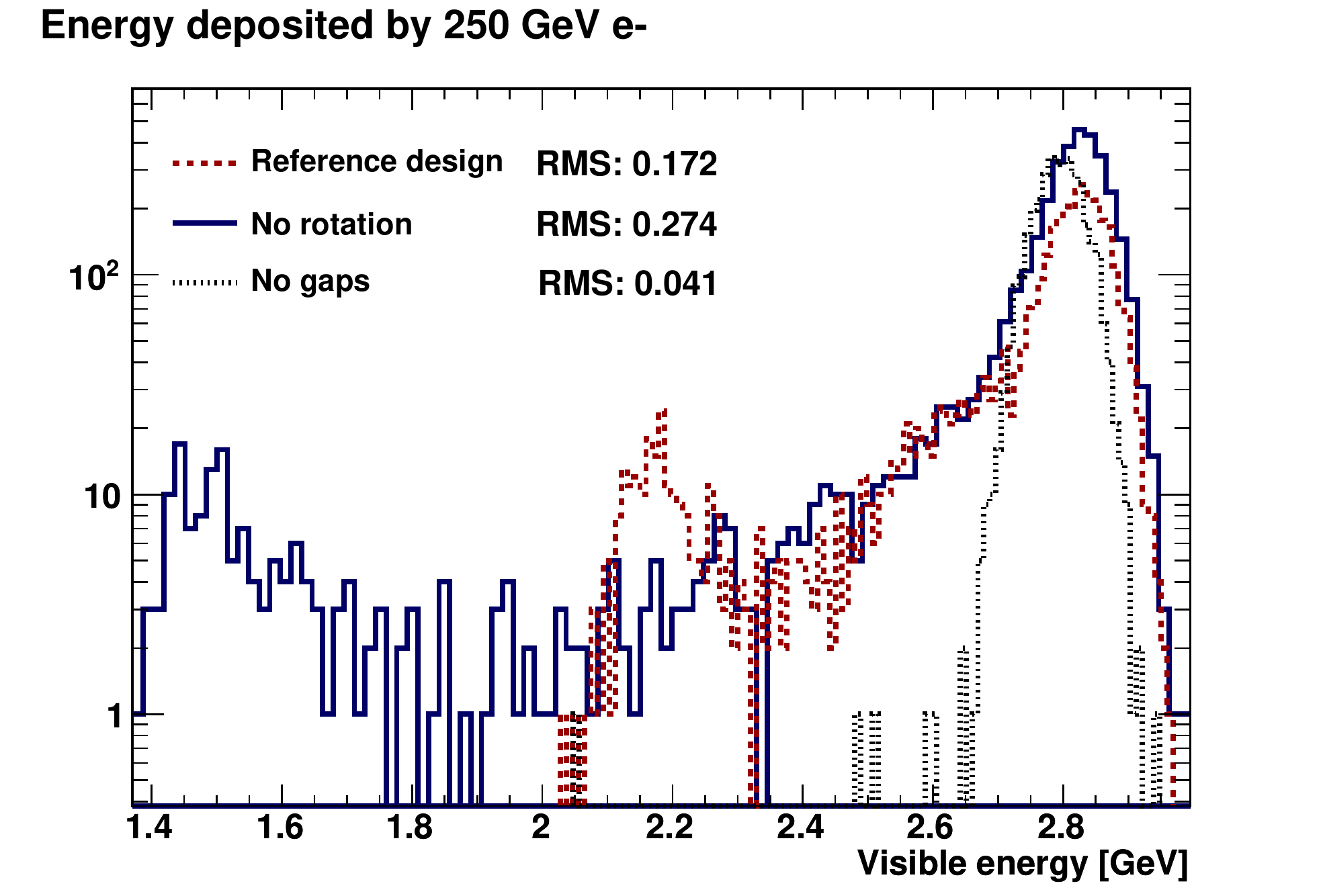}
    \caption{Visible energy at 250 GeV, showing the largest tail for
      the unrotated geometry and no tail for the gapless, ideal
      geometry. The y-axis is shown in log scale to emphasize the
      tails.}
    \label{fig:evis_norot}
  \end{minipage}
  \begin{minipage}[t]{0.5\linewidth}
    \centering
    \includegraphics[width=0.7\textwidth]{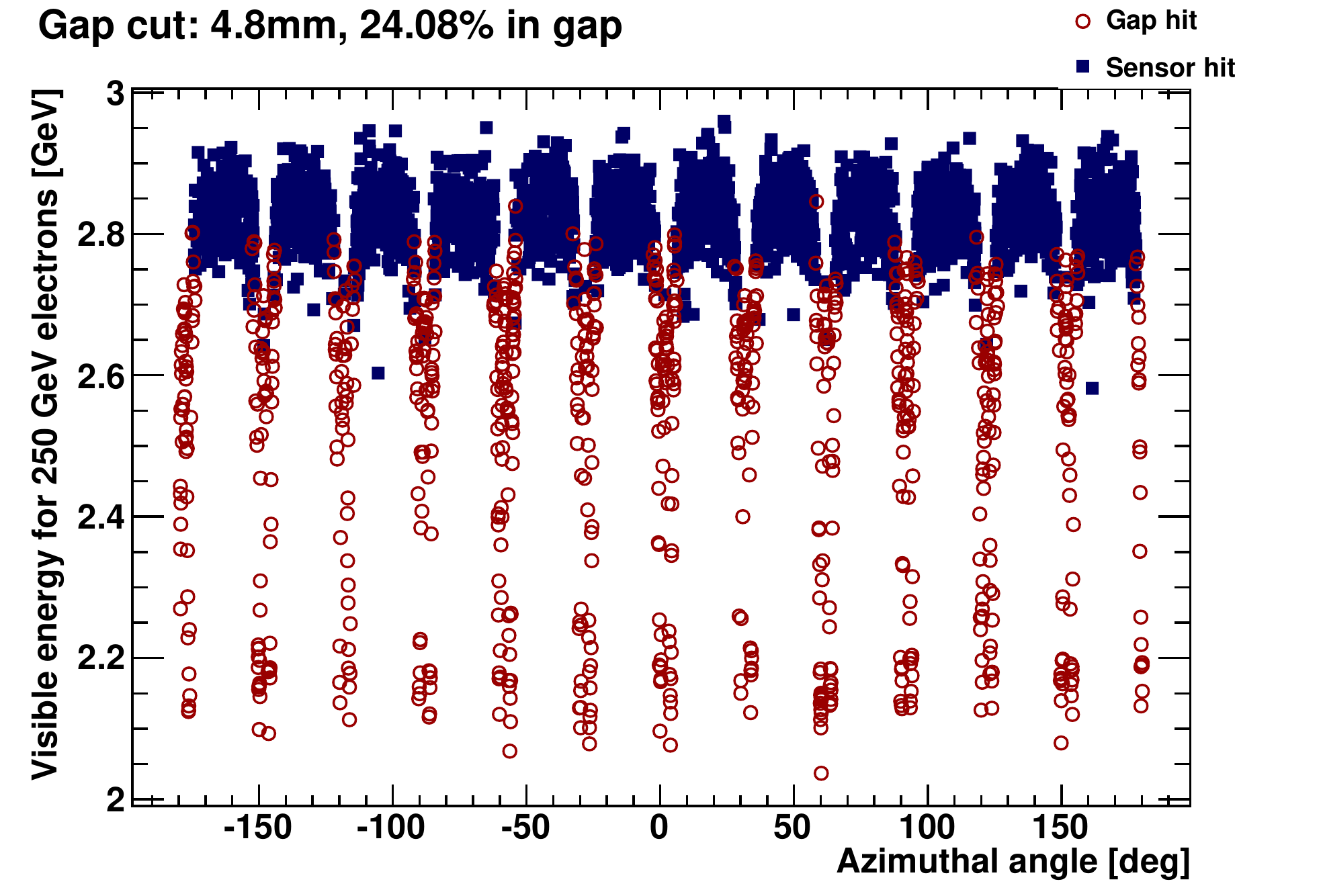}
    \caption{Energy deposition in the tile gaps (gap hits shown in
      red).}
    \label{fig:gap_hits}
  \end{minipage}
\end{figure}

\paragraph{Cuts}
Two methods were investigated for ameliorating the effect of the tile
gaps. The first was simply to discount the energy deposition from primary
electrons incident on the tile gaps. The cut was defined as a distance from
the center of each gap - that is, the energy deposition from all particles
incident within a certain distance from the gaps were ignored. For example,
Figure~\ref{fig:gap_hits} shows the energy deposited by electrons plotted
against the azimuthal angle of the initial point of impact, with a 4.8~mm
selection cut (measured from the center of the tile gap). Only the particles
labeled as ``sensor hits'' -- primary particles which hit the sensors at least the cut width away from a tile gap -- were considered in the calculation of energy resolution. In Figure \ref{fig:gap_hits}, sensor hits are shown as blue squares and gap hits are shown as red circles.

\paragraph{Fits}
The CALICE collaboration showed that for the ILD ECAL, the energy deposition
in similar-style gaps could be fit with a Gaussian distribution
\cite{calice}. Applying similar methods to the LumiCal simulation data, it was
found that the gaps could be fit with a Lorentz distribution and a
Gaussian distribution, allowing for the energy lost in the gaps to be
corrected. The Lorentz distribution fits the sides, and the Gaussian
distribution fits the peak. The equation used for fitting is given in equation~\ref{eq:phi_fitting}:

\begin{equation}
  \label{eq:phi_fitting}
  \mathit{E}(x) = A-\frac{B}{(1+(\frac{x-C}{D})^2)}-E \cdot e^{-F \cdot x^2}
\end{equation}

\begin{figure}[h!]
    \centering
    \includegraphics[width=0.5\textwidth]{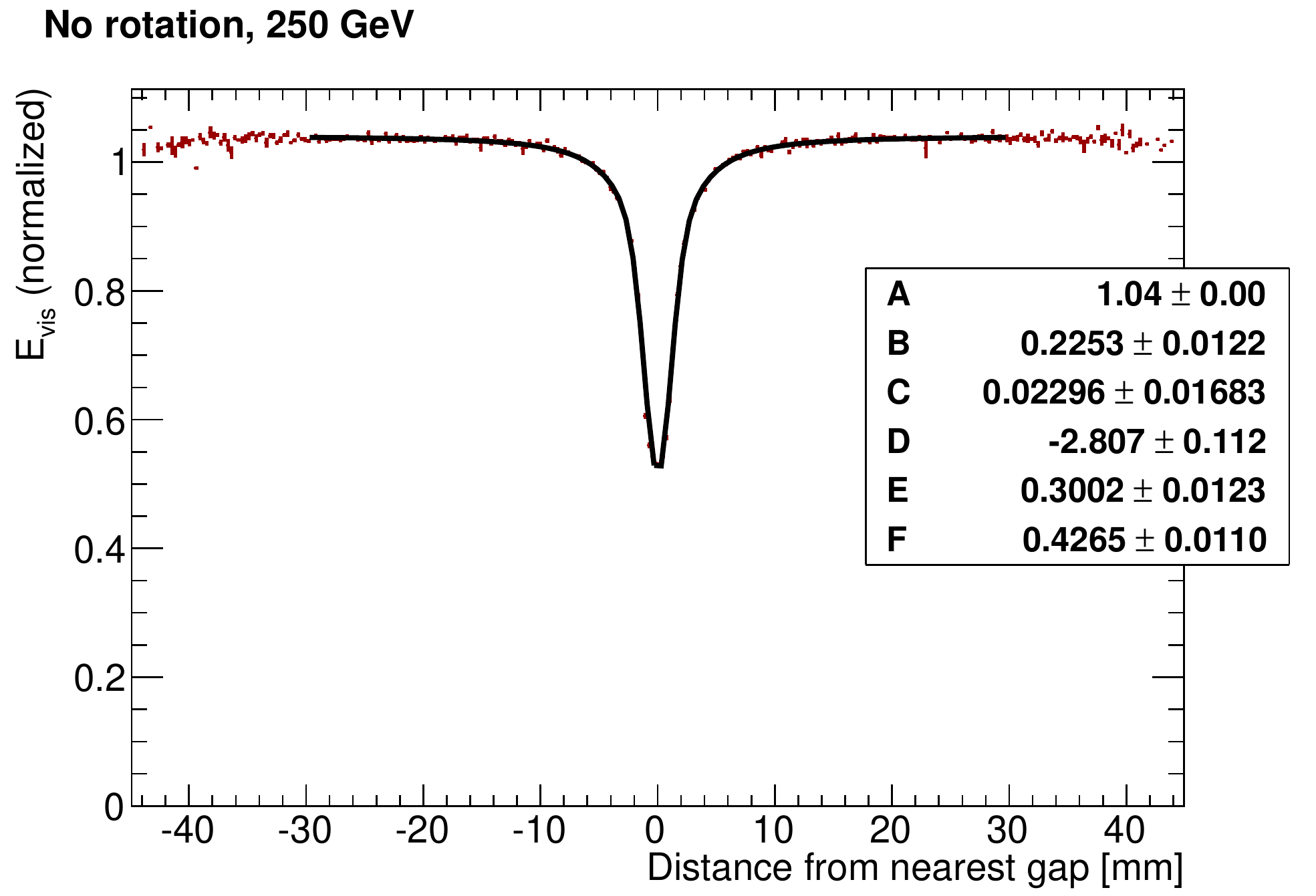}
    \caption{250 GeV energy deposition for all $\phi$ folded onto one
      tile width and normalized to the average deposition in the
      center of each tile.}
    \label{fig:evis_phi_fit}
\end{figure}

Energy deposition was fit against the Cartesian distance of the hit from the
nearest tile gap, instead of azimuthal angle. A fitted plot is shown in Figure~\ref{fig:evis_phi_fit}. The unrotated geometry variant was used for this
analysis as a proof-of-principle. This geometry is nevertheless attractive
from an engineering standpoint because it would reduce the complexity of
construction of the calorimeter.

\begin{figure}[h!]
  \begin{minipage}[t]{0.5\linewidth}
    \centering
    \includegraphics[width=0.7\textwidth]{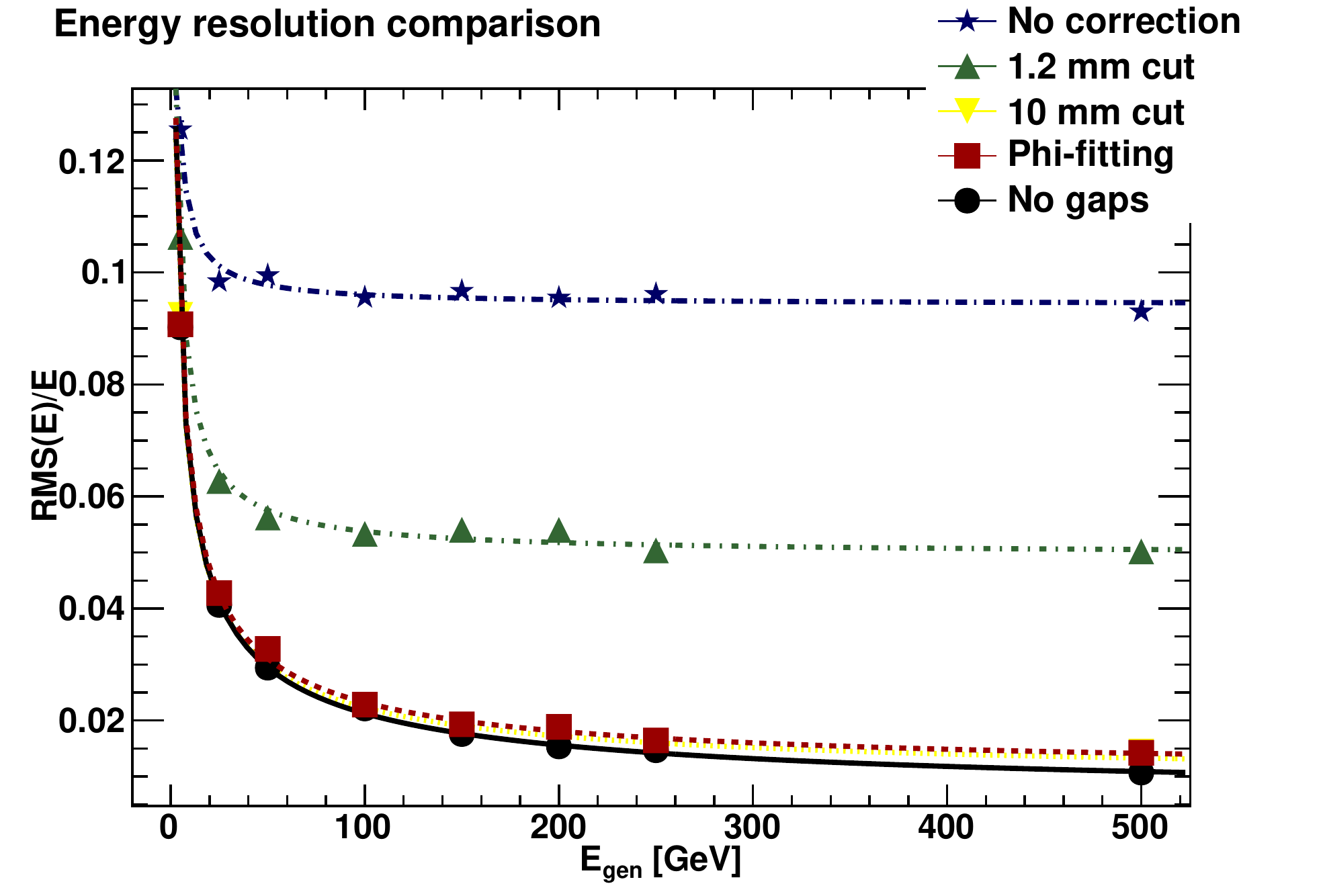}
    \caption{Comparison of different cut widths (width on \emph{each}
      side of the gap) with fit-corrected case and ideal gap-less
      calorimeter case. Error bars are too small to be seen at this
      scale.}
    \label{fig:eres_cutcomp}
  \end{minipage}
  \hspace{0.5cm}
  \begin{minipage}[t]{0.5\linewidth}
    \centering
    \includegraphics[width=0.8\textwidth]{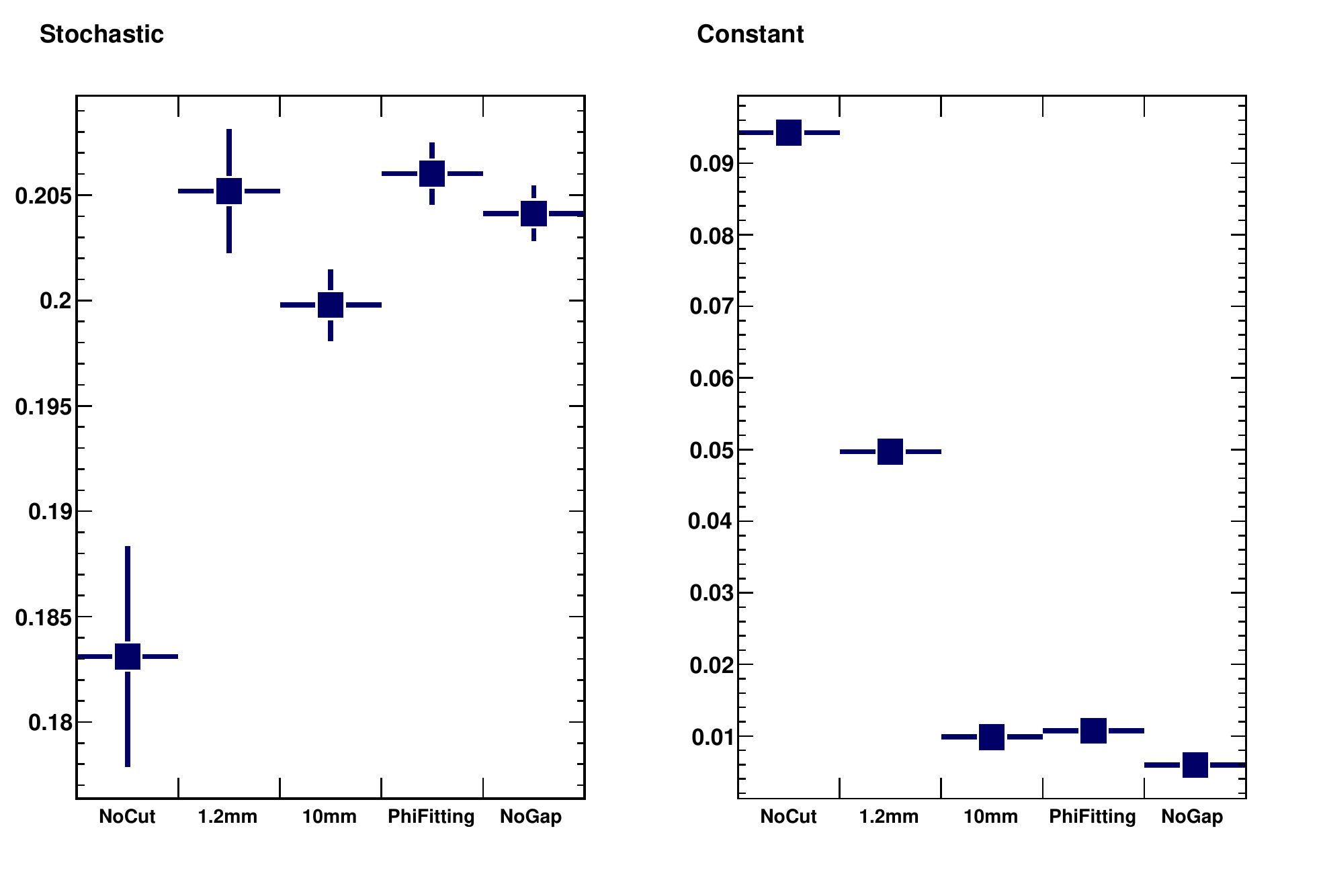}
    \caption{Comparison of fit parameters for different methods of
      determining energy resolution. The stochastic parameter is given on the left, and the constant parameter is given on the right.}
    \label{fig:par_comp}
  \end{minipage}
\end{figure}

Figure~\ref{fig:eres_cutcomp} shows a comparison of the energy resolution performance of the LumiCal variants. The geometry of LumiCal was slightly modified for better comparison with the fitting method - there was no rotation between subsequent layers of LumiCal. The worst performance came from the case without any
correction. The smallest cut, excluding only particles that hit directly in
the tile gaps, already improves the energy resolution by a factor of 2.  The
ideal LumiCal geometry (no gaps) gives the best energy resolution, as
expected. The fitting method approaches the same performance as the ideal
LumiCal design, and is nearly identical to a 10 mm cut (or $\sim$40\% of particles cut). A comparison of fit parameters is shown in Figure~\ref{fig:par_comp}. The uncertainties on the stochastic parameters calculated
from the cutting methods are much larger than those from the $\phi$-fitting
and cases. Furthermore, the constant parameter for $\phi$-fitting almost
completely compensates for leakage, compared to the no-gaps case. This
suggests that performing the $\phi$-fitting correction is the best way to
proceed in making the energy resolution of LumiCal geometrically uniform.

\section{Beam test}
\label{sec:beam_test}

The main purpose of the beam test was to test the performance of the readout
chain from the solid-state sensors (Si or GaAs) up to the front-end ASICs,
which were all custom designed by the FCAL collaboration. The beam test was
performed at DESY-Hamburg. A description of the facilities can be found at
\cite{desy_bt}.

\subsection{Beam area setup}
\label{ssec:beam_area}

\begin{figure}[h!]
  \begin{minipage}[b]{0.5\linewidth}
    \centering
    \includegraphics[width=0.8\textwidth]{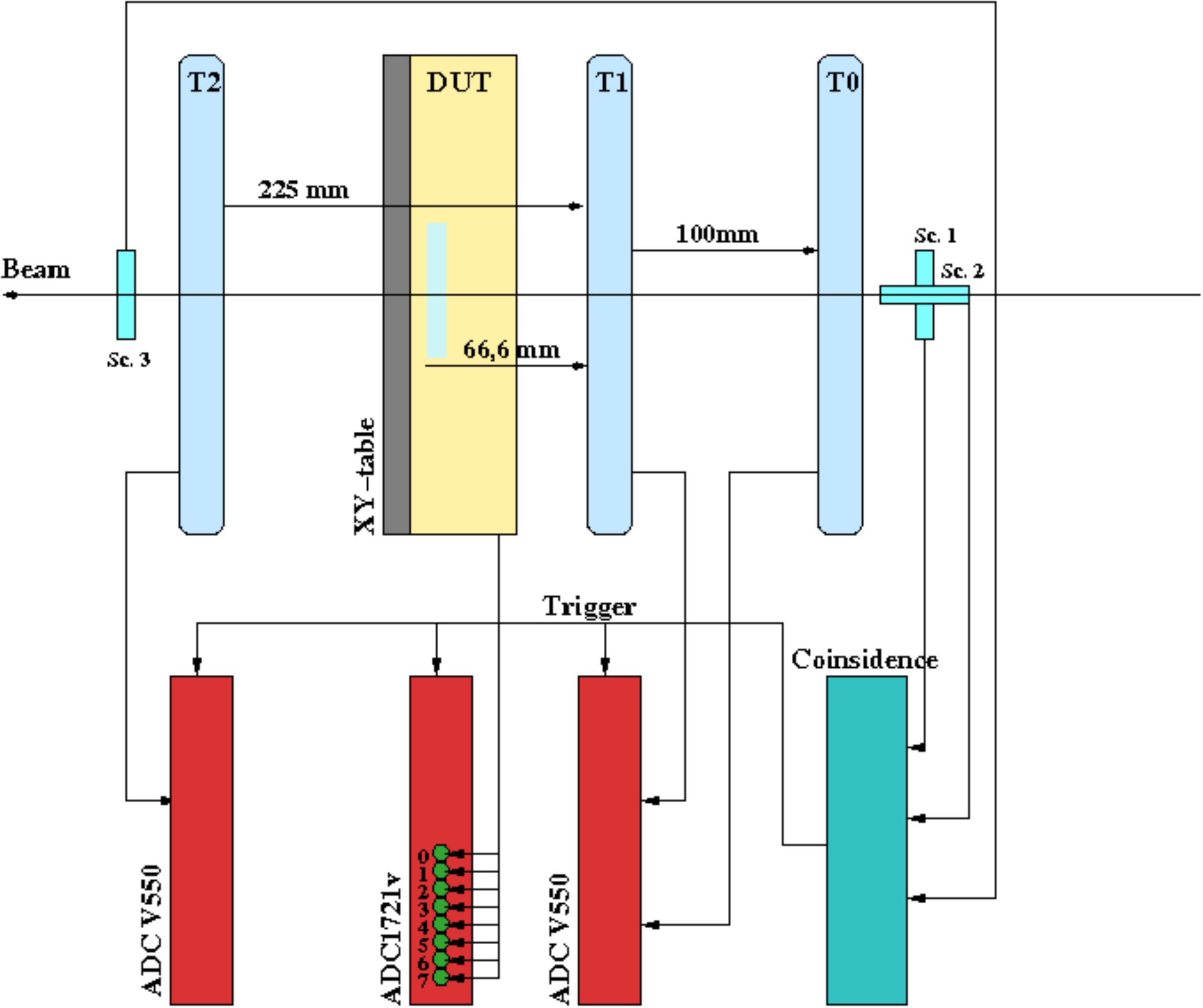}
    \caption{Schematic showing telescope planes and trigger
      electronics, with the DUT in between telescope planes 1 and 2
      (indexed from 0).}
    \label{fig:bt_setup}
  \end{minipage}
  \hspace{0.3cm}
  \begin{minipage}[b]{0.5\linewidth}
    \centering
    \includegraphics[width=0.8\textwidth]{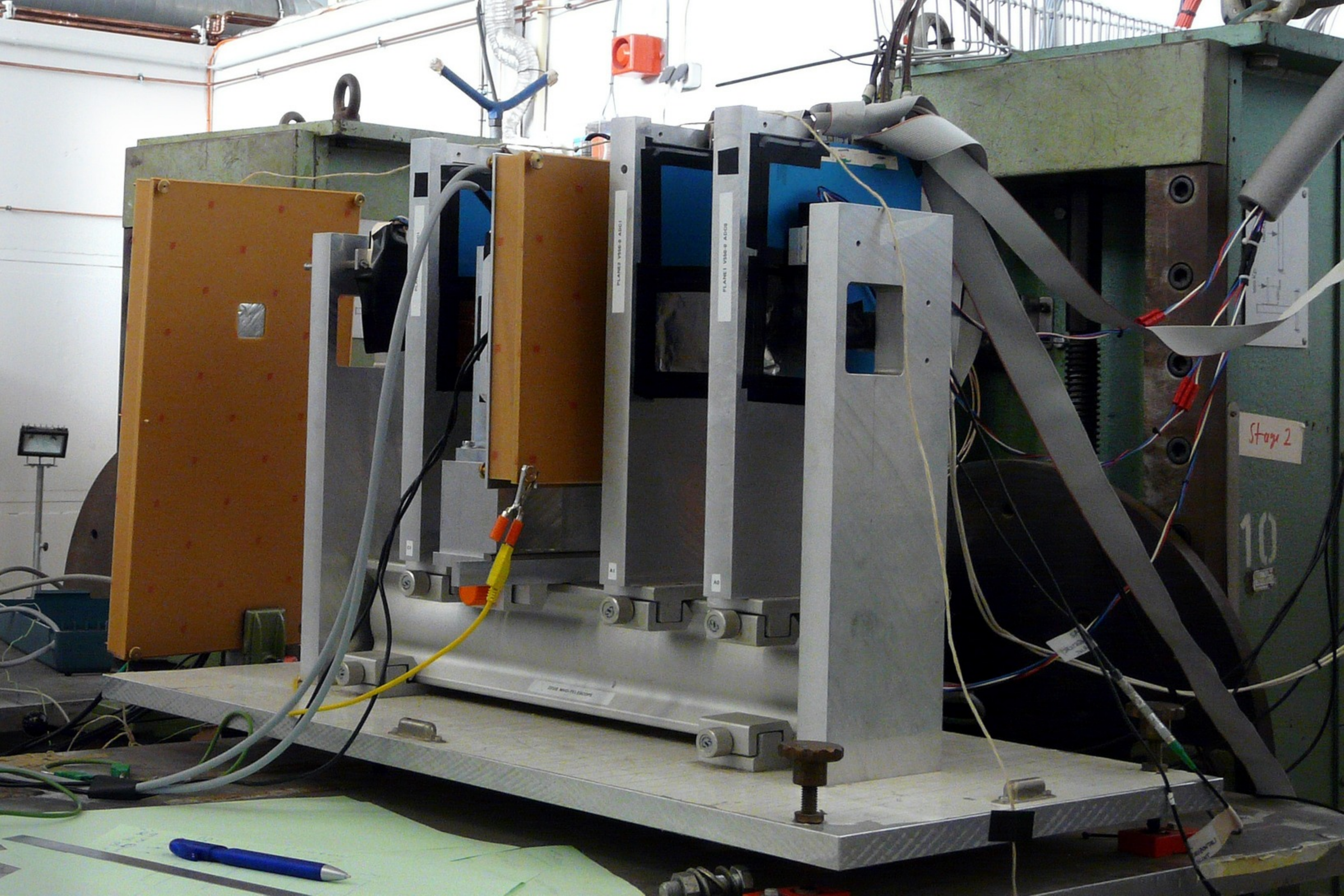}
    \caption{Photograph of BeamCal sensors in place and LumiCal
      sensors behind telescope}
    \label{fig:bt_setup_photo}
  \end{minipage}
\end{figure}

Beamline 22 at DESY contains the ZEUS MVD telescope described in
\cite{mvd_scope}. The telescope, shown schematically in Figure~\ref{fig:bt_setup}, consists of three planes of sensors placed along the beam
axis. Each plane is comprised of two mutually perpendicular sets of Si
microstrips with a 320~$\mu$m pitch. Using the combined data for the three
planes, the track of an electron from the beam line can be determined and the
point of impact on the device under test (DUT) accurately calculated.

Figure~\ref{fig:bt_setup_photo} is a photograph of the setup used for the
BeamCal data runs. The tan DUT box in between planes 1 and 2 contains the
BeamCal readout chain. The tan box after the telescope contains the LumiCal
readout chain, and was placed there for preliminary testing purposes. The
BeamCal box was mounted on a two-dimensional motorized translation stage so
that different parts of the sensor could be irradiated. The LumiCal box was
later moved to the DUT position.

\subsection{Readout chain}
\label{ssec:readout_chain}

\begin{figure}[h!]
  \begin{minipage}[t]{0.5\linewidth}
    \centering
    \subfigure[Complete readout chain: 16 connected sensor pads with
      fanout, FE ASICs bonded to PCB, output buffers, and biasing and
      power blocks.]{\label{fig:readout_chain}\includegraphics[width=0.7\textwidth]{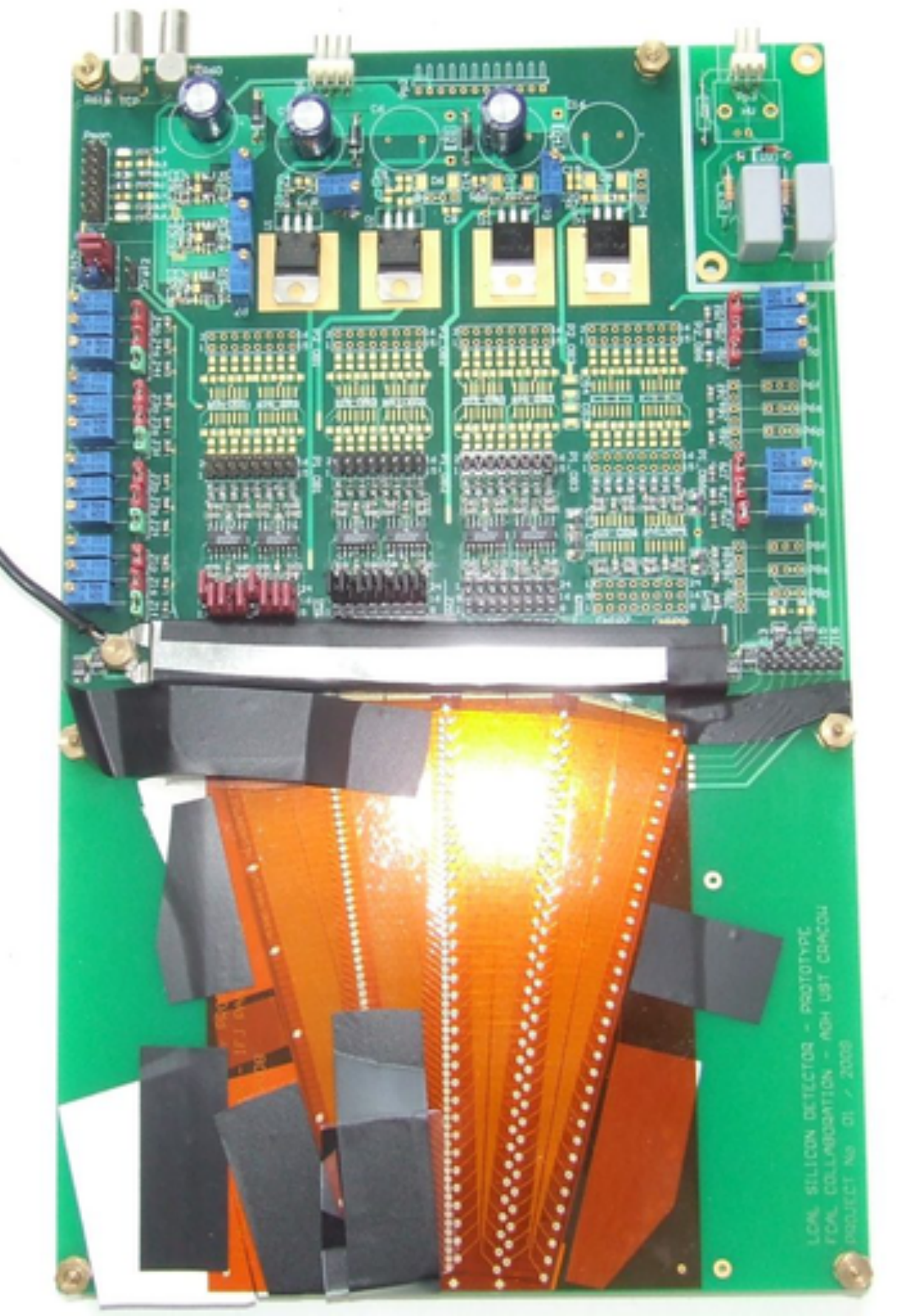}}
  \end{minipage}
  \begin{minipage}[t]{0.5\linewidth}
    \centering
    \subfigure[Picture of the custom FE-ASIC]{ \label{fig:fe_asic_pic}\includegraphics[width=0.7\textwidth]{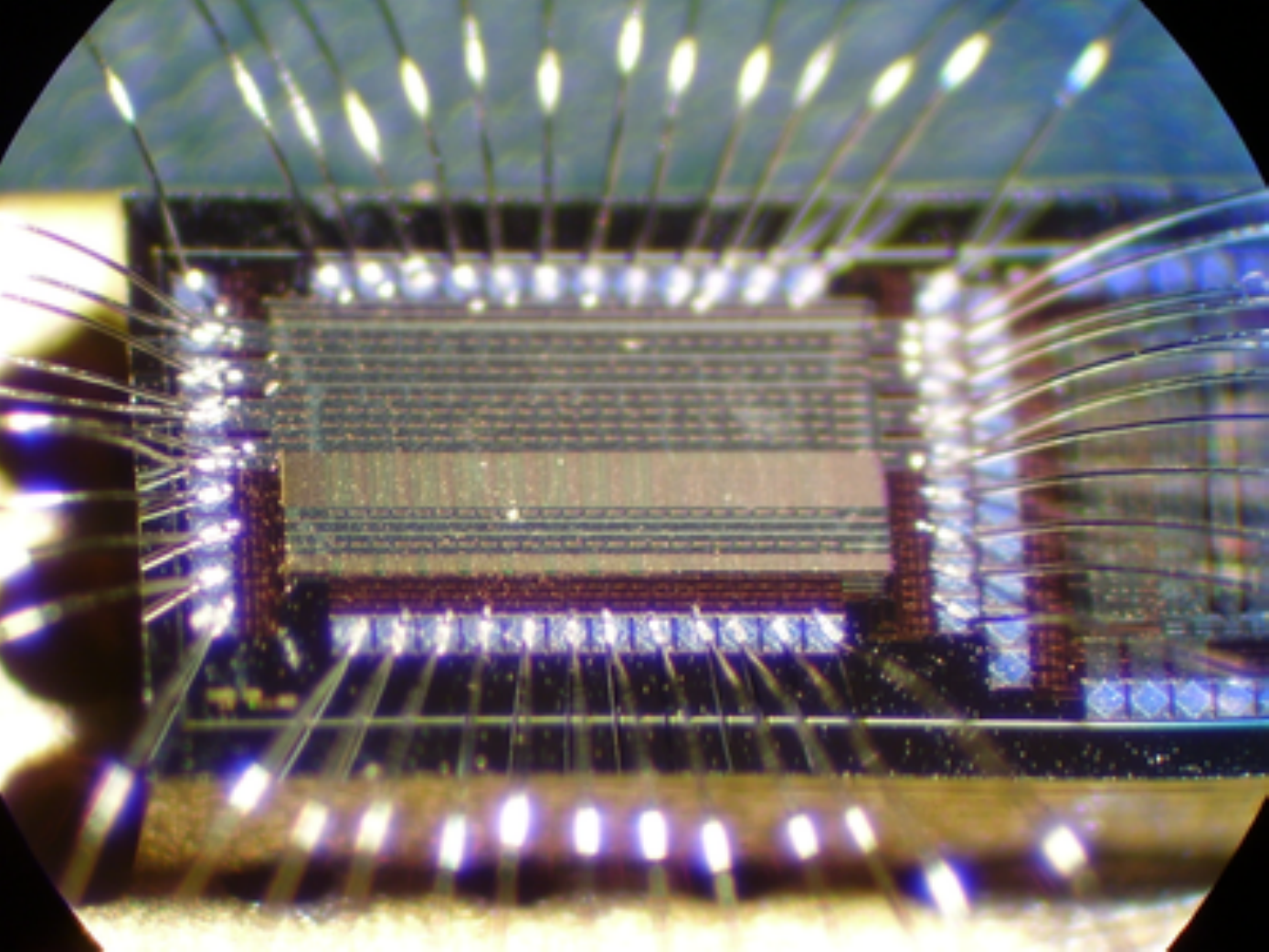}}
    \subfigure[FE ASIC schematic. ``Mode'' switch allows the feedback
    path to be chosen. This schematic shows resistive feedback, but
    an active feedback version was also used.]{\label{fig:fe_asic_schem}\includegraphics[width=0.8\textwidth]{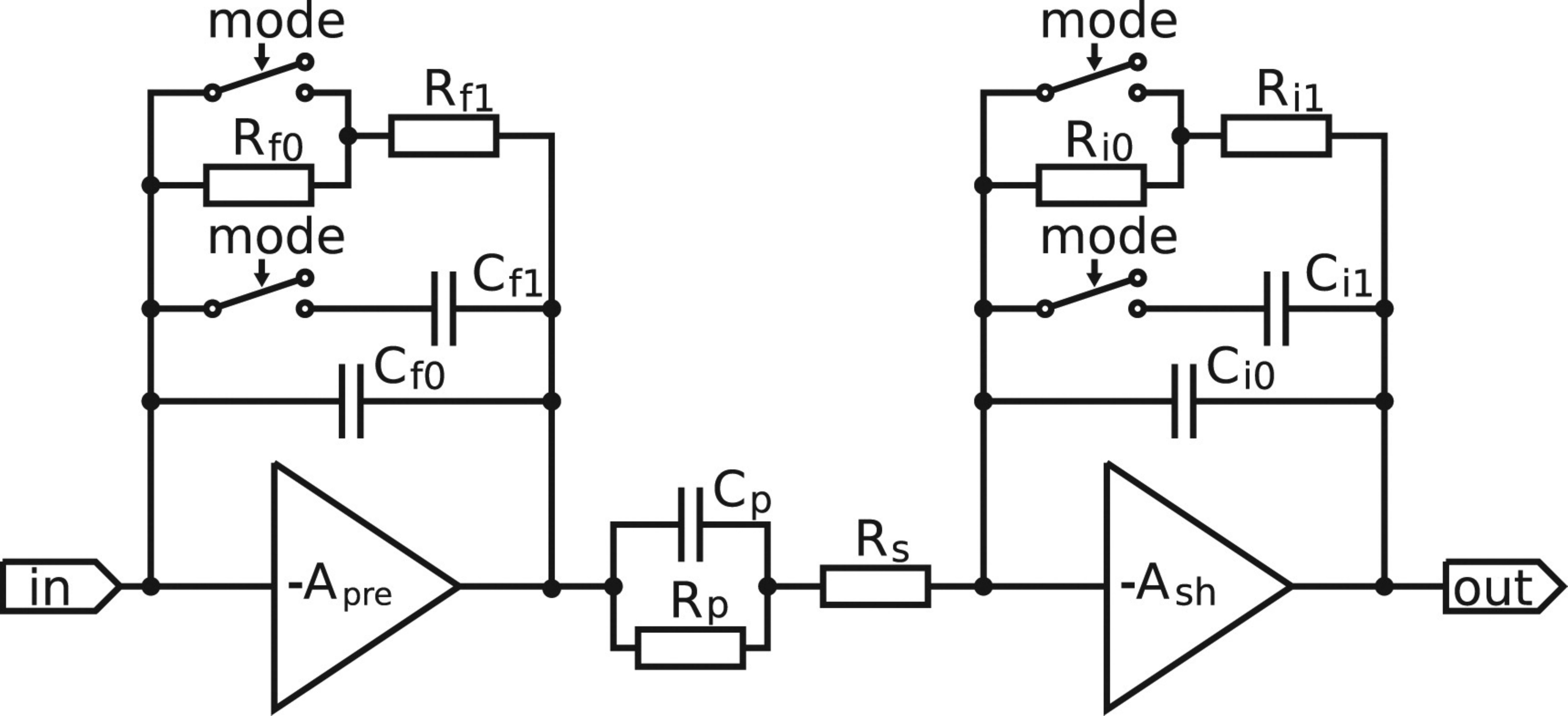}}
    
  \end{minipage}
  \caption{Hardware and electronics under test.}
\end{figure}

Figure~\ref{fig:readout_chain} shows a photograph of the readout chain with
the LumiCal sensors shown in Figure~\ref{fig:lumical_sensors}. The BeamCal
readout chain was identical except for the type of sensors (see section~\ref{ssec:bt_beamcal}). The fanouts were glued to the sensors and connected
to the custom-designed front end ASICs \cite{fe_asic}. The readout chain also
consisted of output buffers, biasing and power blocks, and a commercial ADC
(14-bit, 8-channel Caen VME-1724 for LumiCal; for BeamCal as in Figure~\ref{fig:bt_setup}). In total, sixteen sensor pads were 
connected to the FE ASICs - eight from the bottom of the sensor and eight
from the top. This encompassed the widest range of sensor sizes and fanout
lengths, for crosstalk measurements. For LumiCal, the sensors consisted of one
tile produced by Hamamatsu, with geometry as described in section~\ref{ssec:lumical_geom}. The tile was made from 320~$\mu$m-thick n-doped Si
bulk with p+ pads and Al metallization \cite{sensor_memo}. 

Aside from the GaAs BeamCal sensors (described later), the two eight-channel
front-end ASICs were the most important element of the readout
chain being tested. A picture of the ASIC is shown in Figure~\ref{fig:fe_asic_pic}, and a block diagram is shown in Figure~\ref{fig:fe_asic_schem}.

The FE ASICs feature a mode switch that allows the ASICs to operate in one of
two modes: ``calibration'' mode and ``physics'' mode. The schematic in Figure~\ref{fig:fe_asic_schem} shows the mode switch with passive (resistive)
feedback. For each ASIC, half channels used this type of feedback while the
other half used active feedback. The ASIC design is described in detail
in \cite{fe_asic}, and an analysis of the performance can be found in
\cite{fcal_memo_bt}. 

\subsection{LumiCal results}
\label{ssec:bt_lumical}
\begin{figure}[h!]
  \begin{minipage}[t]{0.5\linewidth}
    \centering
    \includegraphics[width=0.7\textwidth]{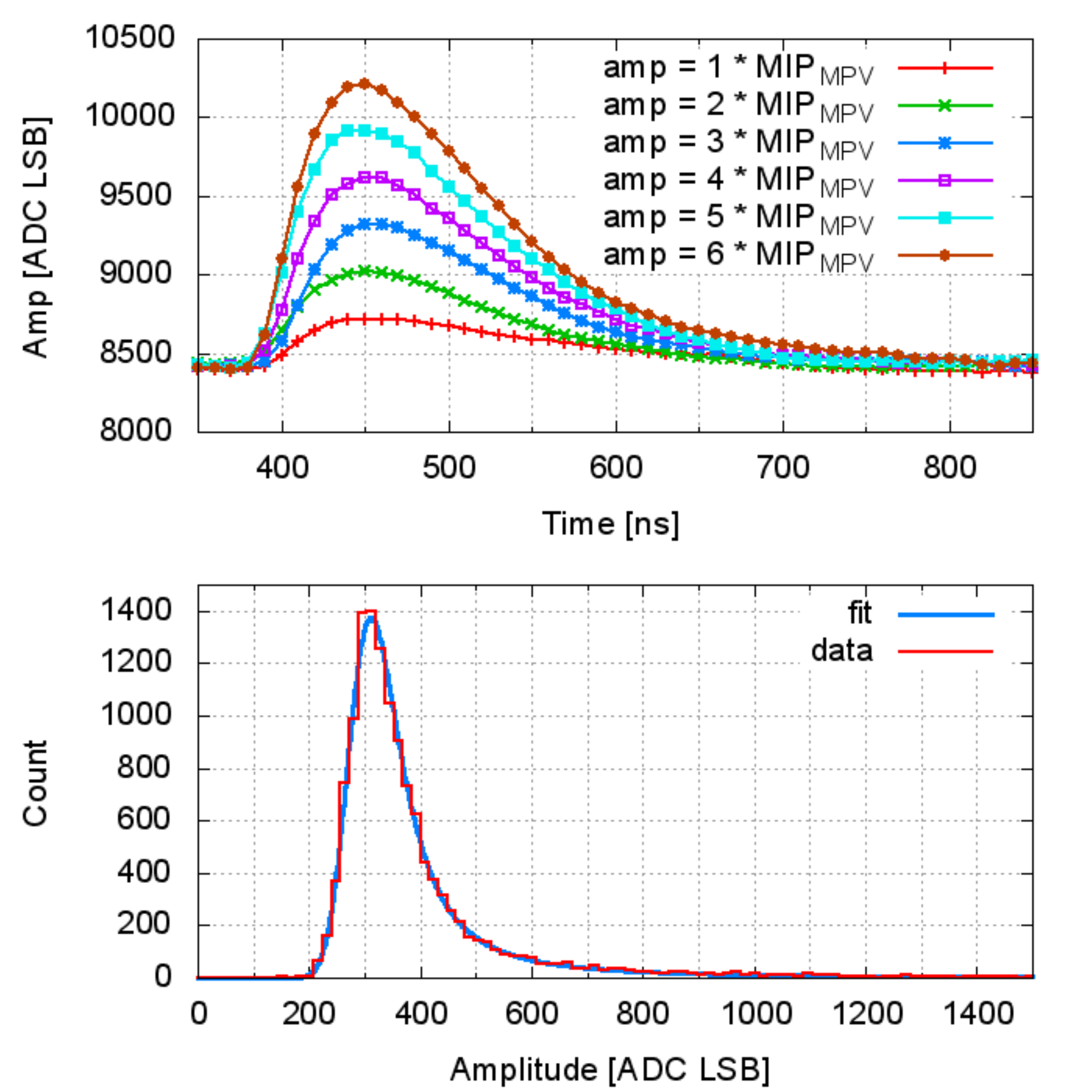}
    \caption{ Top: Response over time to electrons of different energies was uniform. Bottom: Spectrum of signals shows a Landau distribution.}
    \label{fig:bt_landau}
  \end{minipage}
  \begin{minipage}[t]{0.5\linewidth}
    \centering
    \includegraphics[width=0.7\textwidth]{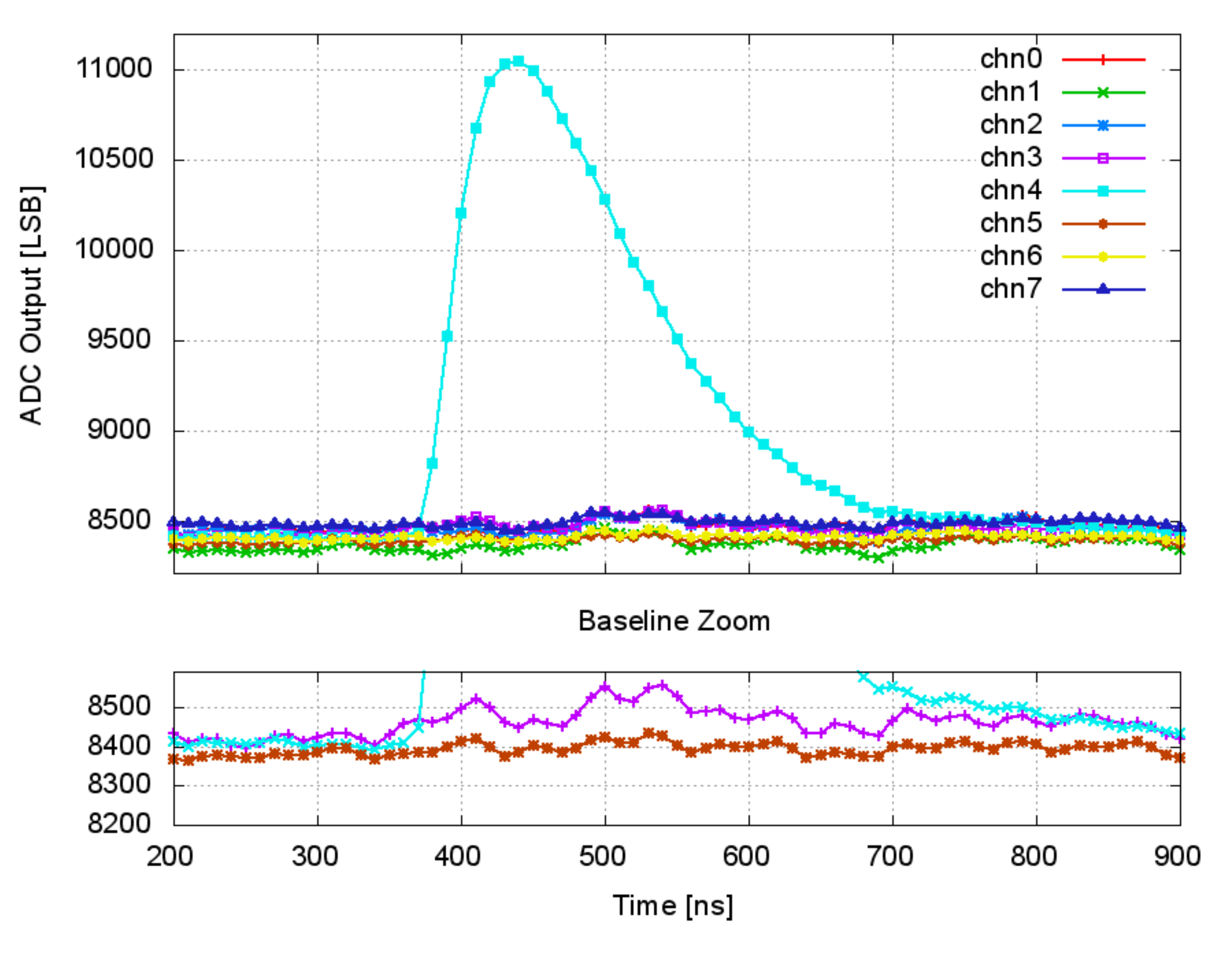}
    \caption{Bottom shows neighboring signals $<$ 1\%}
    \label{fig:bt_crosstalk}
  \end{minipage}
\end{figure}

Preliminary LumiCal beamtest results have been previously reported in an internal FCAL memo \cite{fcal_memo_bt}. Figure~\ref{fig:bt_landau} (top) shows the time
response of a single front-end channel to different amounts of energy
deposition. As expected, the shape of the signal is independent of the signal
amplitude. Figure~\ref{fig:bt_landau} (bottom) shows the energy deposition
spectrum in a single channel for 4.5 GeV electrons from the DESY beamline. The
spectrum for all pads is well fit by a Landau distribution. The set of spectra
were used to find the gain of the channels. The spread of the gain within
each channel type is within 1\%. Signal-to-noise ratio, even for the largest
sensor capacitances, is about 18. 

In multi-channel designs, it is important to ensure good channel-to-channel
separation. Figure \ref{fig:bt_crosstalk} shows the response of eight channels
under test to a particle passing through a sensor pad in the center of the
instrumented area. In channel four, a signal corresponding to approximately 10
MIPS was measured. The lower part of Figure~\ref{fig:bt_crosstalk} shows a
magnified image of the nearest-neighbor baseline signals. The amount of
crosstalk was less than 1\%. 

\subsection{BeamCal results}
\label{ssec:bt_beamcal}

\begin{figure}
  \centering
  \includegraphics[width=0.5\textwidth]{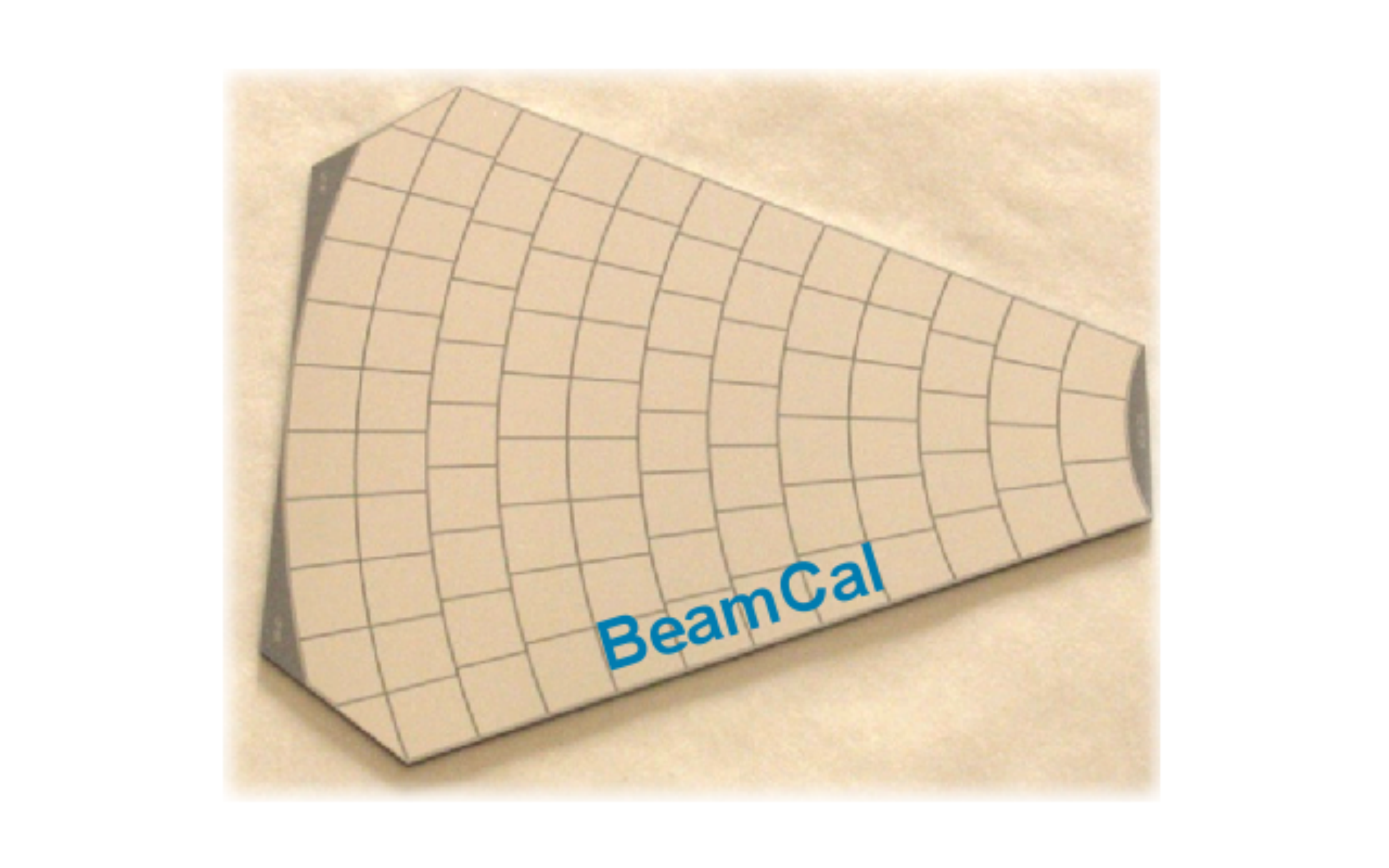}
  \caption{Photograph of GaAs sensor prototype}
  \label{fig:beamcal_sensors}
\end{figure}

The BeamCal sensors, shown in Figure~\ref{fig:beamcal_sensors}, have different
segmentation from the LumiCal sensors shown in Figure~\ref{fig:lumical_sensors}. The square pad size was choosen to be half of the
Moliere radius of electromagnetic showers for high-energy
electrons. Additionally, because BeamCal will operate in a radiation-high
environment, the sensors are proposed to be made of GaAs as a radiation hard
material candidate. GaAs samples were tested in electron beam up to 1.5 MGy \cite{jinst}. The beam test for BeamCal was intended in part to test the
qualities of GaAs for high-energy physics calorimetry. 

\begin{figure}[h!]
  \begin{minipage}[t]{0.5\linewidth}
    \centering
    \includegraphics[width=0.7\textwidth]{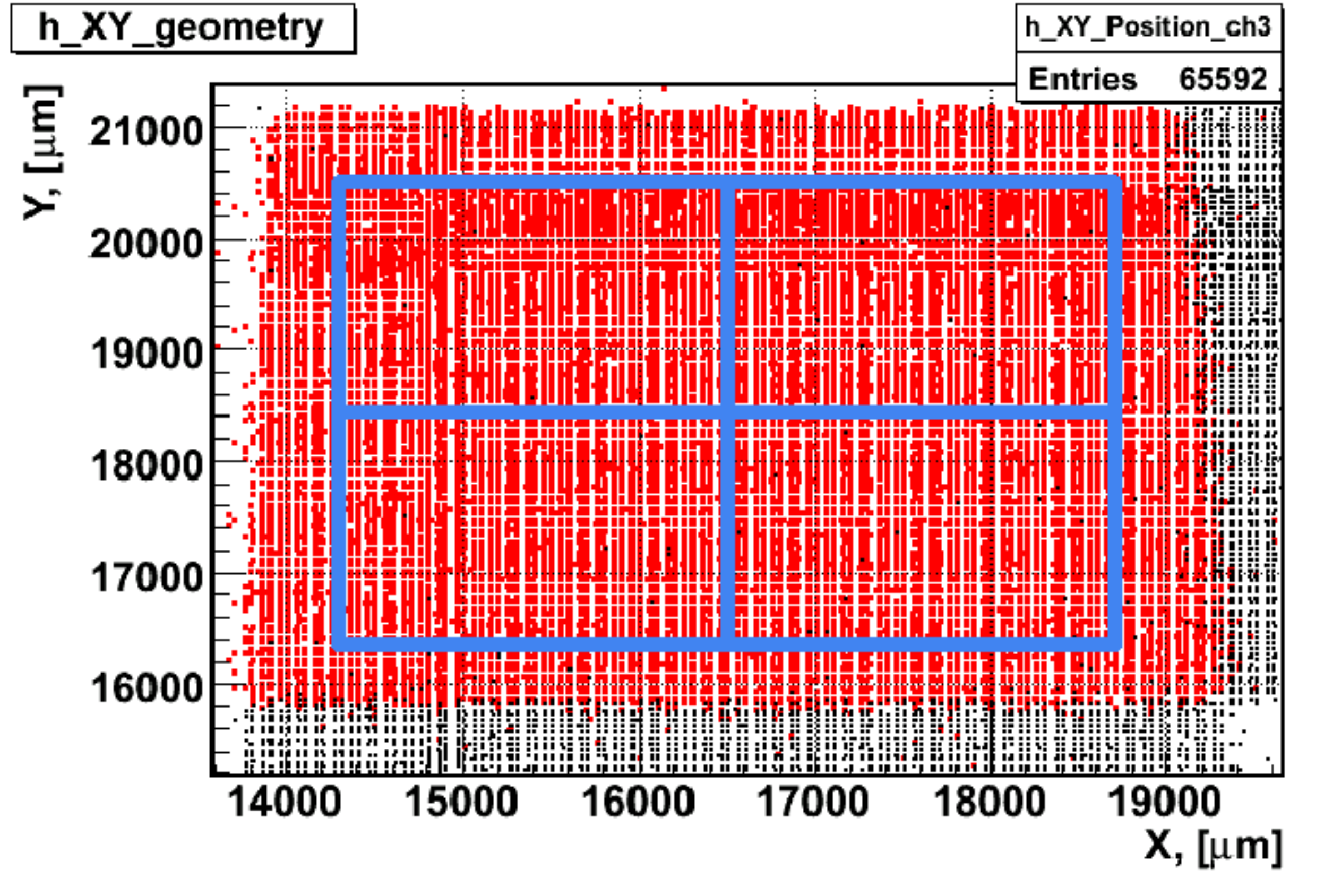}
    \caption{Pad regions in BeamCal.}
    \label{fig:beamcal_pads}
  \end{minipage}
  \begin{minipage}[t]{0.5\linewidth}
    \centering
    \includegraphics[width=0.7\textwidth]{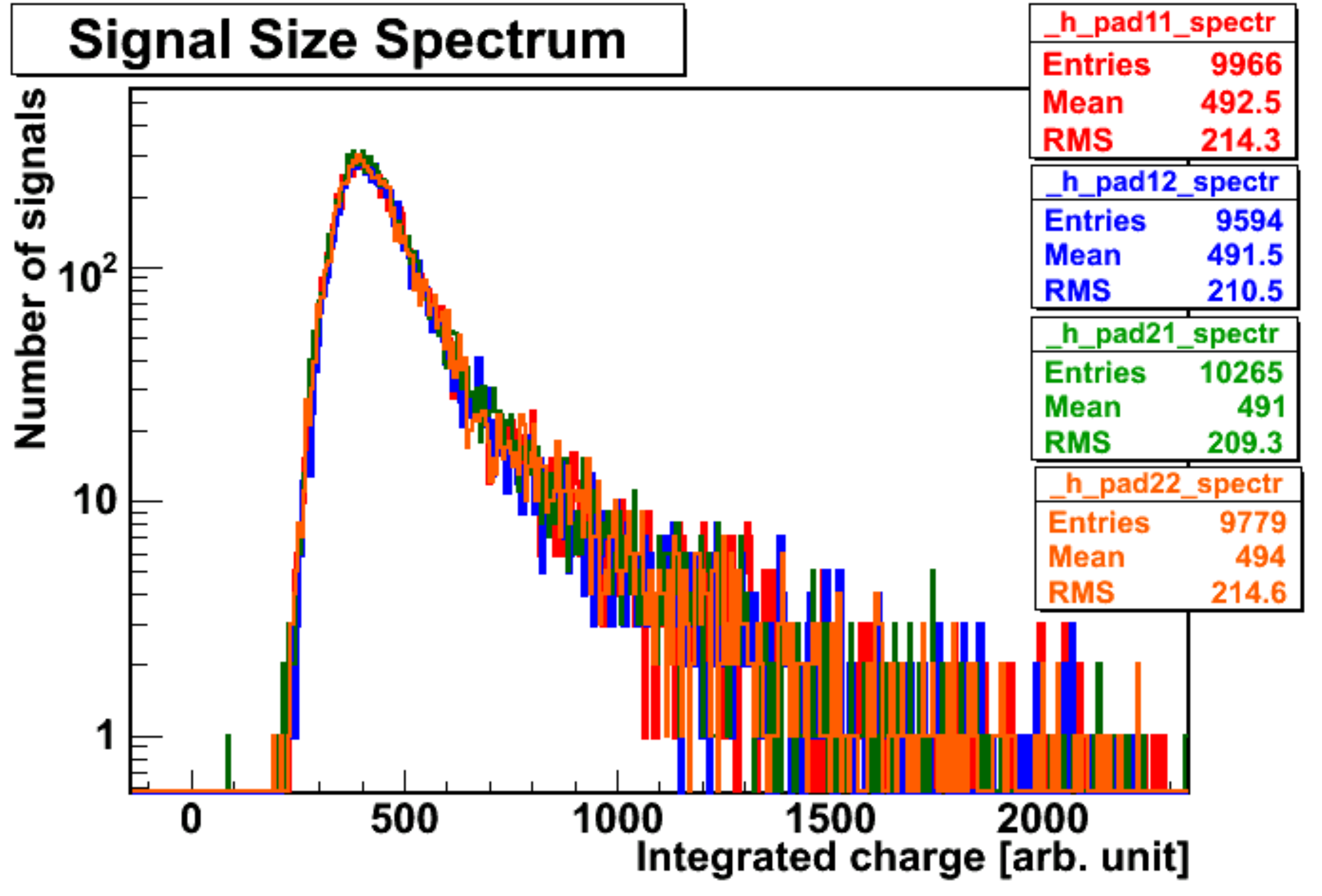}
    \caption{Signals from the four pad areas identified in blue}
    \label{fig:beamcal_signal}
  \end{minipage}
\end{figure}

One important test of the BeamCal sensors was to check for signal uniformity
across each pad. Figure~\ref{fig:beamcal_pads} shows a single 5mm x 5mm pad in red
divided into four regions, indicated by the four squares outlined in blue. In
Figure~\ref{fig:beamcal_signal}, it can be seen that all four regions of the
pad generated identical signal spectra, and furthermore which were all shaped
like a Landau distribution, as expected. The particle detection efficiency was
$\sim$100\%, and SNR was $\sim$20.

The BeamCal beam test results can be found in greater detail in
\cite{olga_proc}.

\section{Conclusions}
\label{sec:conclusions}

The gap-fitting method provides a promising way to recover missing energy
during calculation of energy resolution. This can be done both in the
reference LumiCal geometry and in the unrotated LumiCal geometry, in which
case the performance of the fitting is slightly better. However, should an
unrotated LumiCal be a viable option, further studies must be done to examine
the effect of aligned tile gaps on detector hermeticity.

The beam test proved to be a successful test of the front-end electronics
readout chain developed for both LumiCal and BeamCal. The novel FE ASIC
demonstrated uniform time response to signals of different amplitudes, good
SNR, and had crosstalk of less than 1\% between neighboring channels. BeamCal
GaAs sensors showed good charge uniformity across the sensor pads, supporting
their cause as candidates for solid-state sensors in radiation-hard
environments.

\section{Acknowledgments}
This work was supported by the Commission of the European Communities
under the 7$^{th}$ Framework Programme ``Marie Curie ITN'', grant
agreement number 214560. It was also supported in part by the Polish
Ministry of Science and Higher Education under contract nr 1246/7.PR UE/2010/7.
The authors would also like to thank Cyfronet in Krakow for computing
resources, and Ingrid-Maria Gregor and Artem Kravchenko for their work
at the DESY test beam facility.





\bibliographystyle{elsarticle-num}
\bibliography{tipp2011_aguilar}


\end{document}